\documentclass[12pt]{article}
\usepackage{amsfonts}

\usepackage[normalem]{ulem}

\newcommand{\ex}[1]{{\rm exp}{\left[{#1}\right]}}

\newcommand{\braket}[2]{{\langle {#1}\!\mid\!{#2} \rangle}}
\newcommand{\Hilbert}{{\cal H}}
\newcommand{\field}{{\mathbb F}}

\newtheorem{theorem}{Theorem}[section]
\newtheorem{definition}{Definition}[section]

\newtheorem{corollary}{Corollary}[section]
\newtheorem{property}{Property}[section]
\newtheorem{example}{Example}[section]

\newcommand{\Endproof}{\hfill$\Box$\\}

\usepackage{color}
\def\bR{\begin{color}{red}}
\def\bB{\begin{color}{blue}}
\def\bM{\begin{color}{magenta}}
\def\bC{\begin{color}{cyan}}
\def\bW{\begin{color}{white}}
\def\bBl{\begin{color}{black}}
\def\bG{\begin{color}{green}}
\def\bY{\begin{color}{yellow}}
\def\ec{\end{color}\ }

\usepackage{xy}
\xyoption{matrix}
\xyoption{frame}
\xyoption{arrow}
\xyoption{arc}

\usepackage{ifpdf}
\ifpdf
\else
\PackageWarningNoLine{Qcircuit}{Qcircuit is loading in Postscript mode.  The Xy-pic options ps and dvips will be loaded.  If you wish to use other Postscript drivers for Xy-pic, you must modify the code in Qcircuit.tex}
\xyoption{ps}
\xyoption{dvips}
\fi

\entrymodifiers={!C\entrybox}

\newcommand{\ket}[1]{{\left\vert{#1}\right\rangle}}

\usepackage[margin=3cm,nohead]{geometry}
\begin{document}

\title{On the Concept of Cryptographic Quantum Hashing}

\author{F.  Ablayev\thanks{Kazan Federal University, Kazan, Russian Federation. fablayev@gmail.com} \and M.  Ablayev\thanks{Kazan Federal University, Kazan, Russian Federation}}

\date{September  2015}





%
%
%
%

\maketitle

\begin{abstract}

In the paper we define a notion of quantum resistant ($(\epsilon,\delta)$-resistant) hash function which  combine  together  a notion of   pre-image (one-way) resistance ($\epsilon$-resistance) property we define in the paper  and the notion of  collision resistance ($\delta$-resistance) properties.


We show that in the quantum setting a one-way resistance property and collision resistance property are correlated: the ``more'' a quantum function is one-way resistant  the ``less'' it collision resistant and vice versa.
We present an explicit  quantum hash function which is ``balanced''  one-way resistant and collision  resistant and demonstrate how to build a large family  quantum hash functions. Balanced quantum hash functions need  a high degree of entanglement between the qubits.   We use a ``phase constructions'' technique to express quantum hashing constructions, which  is  good to map hash states  to  coherent states in a superposition of time-bin modes. The later  is ready to be implemented with current optical technology.

\end{abstract}


\section{Introduction}

Quantum cryptography describes the use of quantum mechanical effects  (a) to break cryptographic systems and (b) to perform cryptographic tasks. Quantum factoring algorithm and quantum algorithm for finding discrete logarithm are famous results that belong for the first direction. Quantum key distribution, quantum digital signature schemes constructions belong to the second direction of quantum cryptography.



Gottesman and Chuang proposed in 2001 a quantum digital signature protocol \cite{gc2001} which is based on quantum one-way function. This is also the case for other protocols (see for example \cite{gi2013}).
In \cite{av2014,aa_arxiv2015} we explicitly defined a notion of quantum hashing as a generalization of classical hashing and presented examples of quantum hash functions. It appeared that Gottesman-Chuang  quantum signature schemes are  based on  functions which are actually  quantum hash functions. Those functions have ``unconditionally one-way'' property based on Holevo Theorem \cite{h1973}.  More information on the role of quantum hashing  for the post quantum cryptography, possible application of quantum hashing for quantum signature protocols, and technological expectations for realization of quantum signature schemes are  presented in \cite{k2015}.


  Recall that in the classical setting a cryptographic  hash function $h$ should have the following three properties \cite{aa2015}. (1)
 Pre-image resistance: Given $h(x)$, it should be difficult to find $x$, that is, these hash functions are
one-way functions.
(2) Second pre-image resistance: Given $x_1$, it should be difficult to find an $x_2$, such that $h(x_1) = h(x_2)$.
(3) Collision resistance: It should be difficult to find any  pair of distinct $x_1$, $x_2$, such that $h(x_1) = h(x_2)$. Note, that  there are no one-way functions that are known to be provably more difficult to invert than to
compute, the security of cryptographic hash functions is ``computationally conditional''.


Informally  speaking,  a quantum hash function $\psi$ \cite{av2014,aa_arxiv2015} is a function that maps words (over an alphabet $\Sigma$) of length $k$ to a quantum  pure  states of $s$-qubits ($\psi : \Sigma^k\to ({\cal H}^2)^{\otimes s}$) and has the following properties:
 \begin{enumerate}
 \item  Function $\psi$ must be one-way resistant. In quantum case this means that $k > s$.
 \item   Function $\psi$ must be collision resistant.  In quantum case this means that  for different words $w,w'$ states $\ket{\psi(w)}$, $\ket{\psi(w')}$ must be ``almost orthogonal'' ($\delta$-orthogonal) \cite{aa_arxiv2015}.
 \end{enumerate}
     Quantum collision resistance  property cover  both second pre-image resistance  and collision resistance properties for the quantum setting

In  papers \cite{av2014g,v2015} we considered a quantum Branching Program as a computational model which, we believe, is adequate quantum technological model  for presenting  a quantum communication  protocols and quantum cryptographic signature schemes based on hashing.

\paragraph{Our contribution.}
In the paper we define a notion of $(\epsilon,\delta)$-hash function where values $\epsilon$ and $\delta$ are numerical characteristics of the above two properties: (i) one-way resistance  and (ii)   collision resistance properties. The  notion of the $(\epsilon,\delta)$-hash function is an explicit  generalization of our constructions \cite{av2014,aa_arxiv2015}.
We show that in the quantum setting the one-way resistance property and collision resistance property are correlated: the  ``more'' a quantum function is one-way resistant  the ``less'' it is collision resistant and vice versa.
We present a quantum hash function which is ``balanced''  one-way resistant and collision  resistant. In addition we present more discussion that supports the idea of quantum hashing from our papers.
Note, that a realization of the balanced quantum hash function requires the high degree of entanglement between the qubits which  makes such a state very difficult (or impossible) to create with current technology.

We present  quantum ``balanced'' hashing constructions based on  ``phase transformation''  presentation \cite{a2015}  instead of ``amplitude transformation'' \cite{aa_arxiv2015}.   The phase transformation   is  required  to map quantum hash states  into a sequence of coherent states.  Note, that quantum signature protocols using coherent states can be practically implemented by now day technology that use only a sequence of coherent states, linear optics operations, and measurements with single-photon threshold detectors. See  \cite{al1_2014,al2014,aa2015} for more information and citations.




\section{Quantum  $(\epsilon,\delta)$-Resistant Hash Function}

 Recall  that mathematically a qubit  is described   as a unit vector in the two-dimensional Hilbert complex space ${\cal H}^2$.
Let $s\ge 1$. Let $({\cal H}^2)^{\otimes s}$  be the $2^s$-dimensional  Hilbert space, describing the states of $s$ qubits.
%
%
For an integer $j\in\{0,\dots, 2^s-1\}$ let $\sigma=\sigma_1\dots\sigma_s$ be  a binary presentation of $j$. We use (as usual)  notations $\ket{j}$ and  $\ket{\sigma}$ to denote quantum  state $\ket{\sigma_1}\cdots\ket{\sigma_s}=\ket{\sigma_1}\otimes\dots\otimes\ket{\sigma_s}$.

We let $q$ to be a prime power and $\field_q$ be a finite  field of order $q$. Let $\Sigma^k$  be a set of words of length $k$ over  a finite alphabet $\Sigma$.  Let ${\mathbb X}$ be a finite set. In the paper we  let ${\mathbb X}=\Sigma^k$, or ${\mathbb X}=\field_q$.
For $K=|\mathbb X|$ and integer $s\ge 1$ we define a $(K;s)$ classical-quantum  function (or just quantum function) to be  a unitary transformation  (determined by an element $w\in \mathbb X$) of the initial state $\ket{\psi_0}\in ({\cal H}^2)^{\otimes s}$  to a quantum state $\ket{\psi(w)}\in ({\cal H}^2)^{\otimes s}$


\begin{equation}\label{cqf}
\psi : \{\ket{\psi_0}\}\times \mathbb X \to ({\cal H}^2)^{\otimes s} \qquad \ket{\psi(w)}= U(w)\ket{\psi_0}
\end{equation}
where $U(w)$ is a unitary matrix. We let $\ket{\psi_0}=\ket{0}$ in the paper and  use (for short)  the following  notation (instead the above)
\[ \psi :  \mathbb X \to ({\cal H}^2)^{\otimes s} \quad \mbox{ or } \quad \psi : w\mapsto \ket{\psi(w)}\]

\subsection{One-way Resistant  Function.}

We present the following definition of quantum  $\epsilon$-resistant one-way function. Let ``information extracting''  mechanism ${\cal M}$  be a function ${\cal M} : ({\cal H}^2)^{\otimes s} \to \mathbb X$. Informally speaking   mechanism ${\cal M}$ makes some measurement to state $\ket{\psi}\in ({\cal H}^2)^{\otimes s}$   and decode  the result of measurement to $\mathbb X$.

  \begin{definition}
  Let $X$ be random variable distributed over $\mathbb X$ $\{ Pr[X=w] : w\in{\mathbb X}\}$.    Let $\psi :  \mathbb X \to ({\cal H}^2)^{\otimes s} $ be a quantum function.  Let $Y$ is any random variable  over $\mathbb X$ obtained by some mechanism $\cal M$ making  measurement to  the encoding $\psi$ of $X$  and decoding the result of measurement to $\mathbb X$. Let $\epsilon>0$.
We call a quantum function $\psi$  a one-way $\epsilon$-resistant  function  if    for any mechanism  $\cal M$, the probability $Pr[Y  = X]$ that $\cal M$
successfully decodes $Y$ is bounded by $\epsilon$
\[ Pr[Y  = X] \le \epsilon. \]


  \end{definition}
For the cryptographic purposes  it is natural to expect (and we do this in the rest of the paper) that random variable $X$ is uniformly distributed.

A quantum state of $s\ge 1$ qubits  can ``carry''  an infinite amount of information. On the other hand, fundamental result of quantum informatics  known as Holevo's Theorem \cite{h1973} states that a quantum measurement can only give $s$ bits of information about the state. We will use  here the following  particular  version \cite{n1999} of Holevo's Theorem.

\begin{property}[Holevo-Nayak]\label{h-n}
 Let $X$ be random variable uniformly distributed over a $k$ bit binary words $\{0,1\}^k$. Let $\psi :  \{0,1\}^k \to ({\cal H}^2)^{\otimes s} $ be an $(2^k;s)$ quantum function. Let $Y$ be a random variable  over $\mathbb X$ obtained by some mechanism $\cal M$ making some measurement of the encoding $\psi$ of $X$ and decoding the result of measurement to $\{0,1\}^k$.      Then our probability of correct decoding is given by
\[ Pr[Y=X] \le \frac{2^s}{2^k}.  \]
\end{property}

\subsection{Collision Resistant  Function}
The following definition was presented in \cite{aa_arxiv2015}.
\begin{definition}
\label{QHF}
Let $\delta>0$. We call a quantum function
%
%
 $ \psi : {\mathbb X} \to ({\cal H}^2)^{\otimes s} $
a collision $\delta$-resistant   function if
 for any pair $w,w'$ of different elements,
$$
\left|\braket{\psi(w)}{\psi(w')}\right| \le \delta.  \quad
$$

\end{definition}

\paragraph{Testing Equality.}
What one needs for quantum digital signature schemes realization is an equality testing procedure for quantum hashes  $\ket{\psi(v)}$ and $\ket{\psi(w)}$ in order to compare  classical messages $v$ and $w$; see for example  \cite{gc2001}.
 The  {\em SWAP-test}  is the  known quantum test for the equality of two unknown quantum states $\ket{\psi}$ and $\ket{\psi'}$ (see \cite{gc2001,av2014} for more information).

  We denote $Pr_{swap}[v=w]$ a probability that the {\em SWAP-test} having quantum hashes $\ket{\psi(v)}$ and $\ket{\psi(w)}$ outputs the result ``$v=w$'' (outputs the result ``$\ket{\psi(v)} =\ket{\psi(w)} $'').

 \begin{property}[\cite{gc2001}]\label{swap} Let  function $\psi : w\mapsto \ket{\psi(w)}$ satisfy the following condition.  For any two different elements $v,w\in \mathbb X$ it is true that
$
\left|\braket{\psi(v)}{\psi(w)}\right| \le \delta.
$
Then
\[ Pr_{swap}[v=w] \le \frac{1}{2}(1+ \delta^2). \]
\end{property}
{\em Proof.}  From the description of {\em SWAP}-test  it follows that
$$ Pr_{swap}[v=w] = \frac{1}{2}\left(1+|\braket{\psi(v)}{\psi(w)}|^2\right).$$
\Endproof

The next test for equality was first mentioned in \cite{gc2001}.   We   call this test a \emph{REVERSE-test} \cite{av2014}.
\emph{REVERSE-test} was proposed to check if a quantum state $\ket{\psi}$ is a hash of an element $v$. Essentially the test applies the procedure that inverts the creation of a quantum hash, i.e. it ``uncomputes'' the hash to the initial state.

Formally, let for element $w$ the procedure of quantum hashing   be given by unitary transformation $U(w)$, applied to initial state $\ket{\phi_0}$. Usually we let  $\ket{\phi_0}=\ket{0}$, i.e. $\ket{\psi(w)}=U(w)\ket{0}$. Then the REVERSE-test, given $v$ and $\ket{\psi(w)}$, applies $U^{-1}(v)$ to the state $\ket{\psi(w)}$ and
measures the resulting state with respect to initial state $\ket{0}$. It outputs $v=w$ iff the measurement outcome is $\ket{0}$.  Denote by $Pr_{reverse}[v=w]$ the probability that the \emph{REVERSE-test} having quantum state $\ket{\psi(w)}$ and an element $v$ outputs the result  $v=w$.

\begin{property}\label{reverse} Let hash function $\psi : w\mapsto \ket{\psi(w)}$ satisfy the following condition.  For any two different elements $v,w\in \mathbb X$ it is true that
$
\left|\braket{\psi(v)}{\psi(w)}\right| \le \delta.
$
Then
\[ Pr_{reverse}[v=w] \le \delta^2. \]
\end{property}
{\em Proof.} Using the property that unitary transformation keeps scalar product we have that
$$
 Pr_{reverse}[v=w]  = |\braket{0}{U^{-1}(v)\psi(w)}|^2 = |\braket{U^{-1}(v)\psi(v)}{U^{-1}(v)\psi(w)}|^2 =
 |\braket{\psi(v)}{\psi(w)}|^2 \le \delta^2.
$$
\Endproof
\subsection{One-way Resistance and Collision Resistance}
The above two definitions and considerations lead to the following formalization of the quantum cryptographic (one-way  and collision resistant) function
\begin{definition}\label{main_def}
Let $K=|\mathbb X|$ and $s\ge 1$. Let $\epsilon >0$ and $\delta>0$.  We call a function
 $ \psi : {\mathbb X} \to ({\cal H}^2)^{\otimes s}
 $
a quantum $(\epsilon,\delta)$-Resistant $(K;s)$-hash function iff $\psi$
 is a one-way $\epsilon$-resistant  and
 is a collision $\delta$-resistant function.
\end{definition}

We present below the following two examples to demonstrate  how one-way $\epsilon$-resistance and collision $\delta$-resistance are correlated. The first example was presented in \cite{af1998} in terms of quantum automata.


\begin{example}\label{enc1}
 Let us encode numbers $v$ from $\{0,\dots ,2^k-1\}$  by a single qubit as follows:
\[ \psi : v \mapsto  \cos\left(\frac{2\pi v}{2^k}\right)\ket{0} +
  \sin\left(\frac{2\pi v}{2^k}\right)\ket{1}. \]
%
\end{example}
%
%
Extracting  information from $\ket{\psi}$ by measuring $\ket{\psi}$ with respect to the basis $\{\ket{0},\ket{1}\}$ gives the following result. The function $\psi$ is one-way $\frac{1}{2^k}$-resistant (see Property \ref{h-n}) and  collision $\cos\left(\pi/ {2^{k-1}}\right)$-resistant.  According to the properties \ref{h-n} and \ref{reverse} the function $\psi$ has good one-way property, but has bad resistance property for a large $k$.

%

\begin{example}\label{enc2}
  We consider a number  $v\in \{0,\dots ,2^k-1\}$ to be also a binary word $v\in\{0,1\}^k$. Let $v=\sigma_1\dots \sigma_k$.  We  encode $v$ by $k$ qubits:
   $ \psi : v \mapsto  \ket{v}= \ket{\sigma_1}\cdots\ket{\sigma_k}$.
\end{example}
Extracting  information from $\ket{\psi}$ by measuring $\ket{\psi}$ with respect to the basis $\{\ket{0\dots 0}, \dots, \ket{1\dots 1}\}$ gives the following result. The function $\psi$ is one-way $1$-resistant and collision $0$-resistant. So, in contrary to the   Example \ref{enc1}  the encoding  $\psi$ from the Example  \ref{enc2} is   collision free, that is,  for different words $v$ and $w$ quantum states $\ket{\psi(v)}$ and  $\ket{\psi(v)}$ are orthogonal and therefore  reliably distinguished; but we loose the one-way property:  $\psi$   is easily invertible.

The following result \cite{aa_arxiv2015} shows that quantum  collision $\delta$-resistant $(K;s)$ function  needs at least $\log\log{K} - c(\delta)$ qubits.

%
\begin{property}[\cite{aa_arxiv2015}] \label{lb} Let $s\ge 1$ and $K=|\mathbb X|\ge 4$. Let
$ \psi : \mathbb X \to ({\cal H}^2)^{\otimes s}$
be a  $\delta$-resistant $(K;s)$ hash function. Then
\[s \ge   \log\log {K} - \log\log\left(1+\sqrt{2/(1-\delta)}\right) -1.\]

\end{property}

Properties \ref{lb} and \ref{h-n} provide a basis for building a
 ``balanced'' one-way $\epsilon$-resistance and collision $\delta$-resistance properties.  That is, roughly speaking, if we need to hash  elements $w$ from a domain $\mathbb X$ with $|\mathbb X|=K$ and if  one can build for a $\delta>0$ a collision  $\delta$-resistant $(K;s)$ hash function $\psi$ with $s \approx   \log\log{K}-c(\delta)$ qubits then the function $f$ will be a one-way $\epsilon$-resistant with  $\epsilon \approx ({\log{K}}/{K})$.



\section{``Balanced'' Quantum Hash Functions Constructions}
We start by  recalling  some definitions, notations, and facts from \cite{rsw1993}.  For a field $\field_q$, the discrete Fourier transform of  a set $B\subseteq \field_q$ is the function
\[ f_B(w)=\sum_{b\in B}\ex{i\frac{2\pi wb}{q}} \]
defined for every $w\in\field_q$. let $\lambda(B)=\max_{w\not= 0}|f_B(w)|/|B|$.
 For $\delta >0$ we define $B\subseteq \field_q$ to be $\delta$-good if $\lambda(B)\le\delta$.  By $B_{\delta,q}$ we denote $\delta$-good subset of $\field_q$.
For a field $\field_q$, let $B\subseteq \field_q$. For every $b\in B$ and $w\in\field_q$, define a function $h_b :\field_q \to \field_q$  and a family $H_B$ by the rule
  \[   h_b(w)=bw\pmod{q}, \qquad H_B=\{h_b: b\in B\}.\]
   We denote   by $H_{\delta,q}$ the above set of functions and call $H_{\delta,q}$  $\delta$-good
    if $B=B_{\delta,q}$ is  $\delta$-good.


\begin{theorem}\label{qhg-th}
Let $\delta>0$ and $q$ be a prime power. Let $H_{\delta,q}=\{h_1,\dots, h_T\}$ be  $\delta$-good. Then for $s=\log T$  a function
\begin{equation}\label{good-h}
 \ket{\psi_{H_{\delta,q}}(w)} =\frac{1}{\sqrt{T}}\sum_{j=1}^T \ex{i\frac{2\pi h_j(w)}{q}}\ket{j}.
\end{equation}
 is a collision  $\delta$-resistant  $(q; s)$ quantum hash function.
\end{theorem}
{\em Proof.} Note, that the proof of this theorem in terms of amplitude transformation was presented in \cite{av2014}. The proof presented below is in terms of phase transformation \cite{a2015}.

Let $B_{\delta,q}=\{b_1,\dots, b_T\}$ determines $\delta$-good family $H_{\delta,q}$. We let  $H=H_{\delta,q}$   in the proof.
We consider the following  quantum function   $\psi_{H} :\field_q\to ({\cal H}^2)^{\otimes s}$  of function $\psi_H$
\[ \ket{\psi_{H}(w)} =
\frac{1}{\sqrt{T}}\sum_{j=1}^T \ex{i\frac{2\pi h_j(w)}{q}}\ket{j} =
\frac{1}{\sqrt{T}}\sum_{j=1 }^T \ex{i\frac{2\pi wb_j}{q}}\ket{j}. \]
The quantum state $\ket{\psi_{H}(w)}$ composed from $s$ qubits. To show that $\psi_{H}$ is collision  $\delta$-resistant $(q; s)$ quantum hash function
we prove the collision $\delta$-resistance of $\psi_{H}$. Consider a pair $w,w'$ of different elements from $\field_q$ and their inner product $\braket{\psi_{H}(w)}{\psi_{H}(w')}$. Recall that the inner  product  of two complex vectors $\ket{\alpha}=(\alpha_1,\dots, \alpha_T)$ and $\ket{\beta}=(\beta_1,\dots, \beta_T)$ is the sum
$ \braket{\alpha}{\beta}= \sum_j \alpha_j\bar{\beta}_j$
 where $\bar{\beta}_j$ is the complex conjugate of $\beta_j$.
Using the fact that  the conjugate of  $e^{i \phi}$ is  $e^{-i \phi}$,  and  the fact that $B_{\delta,q}$  is $\delta$-good we have that
 \[ \braket{\psi_{H}(w)}{\psi_{H}(w')} =
  \frac{1}{T}\sum_{b\in B_{\delta,q}} \ex{i\frac{2\pi (w-w')b}{q}}
  \le \lambda(B_{\delta,q})\le  \delta . \]
 \Endproof
\begin{itemize}
\item In \cite{aa_arxiv2015} we defined a set of discrete functions  a {\em quantum hash generator}  if it allow to built a  quantum hash function.
 \end{itemize}
In the context of Theorem \ref{qhg-th}  the set  $H_{\delta,q}$ is a collision $\delta$-resistant   hash generator: it generates the   quantum hash function $\psi_{H_{\delta,q}}$.

\paragraph{Optimality of the hashing scheme.}
The following facts were presented in \cite{rsw1993}.
%
Let $\delta=\delta(q)$ be any function tending to zero as $q$ grows to infinity. Then there exists $\delta$-good set  $B_{\delta,q}$ with $|B_{\delta,q}|=(\log{q}/\delta(q))^{O(1)}$.
%
Several optimal (in the sense of the above lower bound) explicit constructions of $\delta$-good sets $B_{\delta,q}$ were presented by different authors. For those constructions
    \[
    \delta(q)=\frac{1}{(\log{q})^{O(1)}}
  \quad  \mbox{ and } \quad
    |B_{\delta,q}|=(\log{q})^{O(1)}.
    \]

The following statement summarize Theorem \ref{qhg-th} and the  above consideration.

\begin{corollary}\label{good_H} Let $q$ be a prime power, $T(q)=(\log{q})^{O(1)}$, and $s=\log{T(q)}$. Let $\epsilon(q) = {T(q)}/{q}$ and $\delta(q)={1}/T(q)$. Let $H_{\delta,q}$ be $\delta(q)$-good set of functions with $|H_{\delta,q}|=T(q)$.
 Then
 \begin{enumerate}
\item  $\psi_{H_{\delta,q}}$ is ``balanced'' quantum $\left(\epsilon(q), \delta(q)\right)$-resistant quantum $(T(q);s)$-hash function.

\item The  number $s$  of qubits  is good in the sense of the lower bound of Property  \ref{lb} which gives the following lower bound
$ s\ge \log{\log{q}}-\log{\log{\left( 1+\sqrt{2/\delta}\right)}}-1$.
\end{enumerate}
\end{corollary}

We refer to the paper \cite{av2014} for more information on practical construction of the set $H_{\delta,q}$ and the Numerical results from genetic algorithm for $H_{\delta,q}$ construction.

\paragraph{Balanced  Quantum Hash Function Families. }
In \cite{aa_arxiv2015} we   offered design, which allows to build a large amount of different quantum hash functions.  The construction is based on composition of classical $\epsilon$-universal hash family  \cite{st1996} and a given family $H_{\delta,q}$ a quantum hash generator. A  resulting family of functions is a new quantum hash generator. In particular, we present a quantum hash generator $G_{RS}$ based on Reed-Solomon code.

%
%
%
%

Let $q$ be a prime power, let $k\le n\le q$, let $\field_q$ be a finite field. A {\em  Reed-Solomon} code (for short RS-code) is a linear code
$ C_{RS}:(\field_q)^k\to (\field_q)^n $
 defined as follows. Each word $w\in(\field_q)^k$, $w=w_0w_1\dots w_{k-1}$ associated with the polynomial
$P_w(x)= \sum_{i=0}^{k-1}w_ix^i$.
 Pick $n$ distinct elements (evaluation points) $A=\{a_1,\dots, a_{n}\}$  of $\field_q$. A common special case is $n=q-1$ with the set of evaluating points being $A=\field_q\backslash\{0\}$. To encode word $w$ we evaluate $P_w(x)$ on over  all $n$ elements $a\in A$
$  C_{RS}(w)=(P_w(a_1)\dots P_w(a_{n}))$.

We define  family  $F_{RS}=\{ f_a :  a\in A\}$ based on RS-code $C_{RS}$ as follows. For $a\in A$ define $f_a: (\field_q)^k \to \field_q$ by the rule
$ f_a(w)= P_w(a)$.
Let $H_{\delta, q}= \{h_1,\dots, h_T\}$ be a $\delta$-good set of functions, satisfying Corollary \ref{good_H}.
Composition
$$G_{RS}=F_{RS}\circ H_{\delta, q}=\{ g_{jl} : g=h_j(f_{a_l}), h_j\in H_{\delta, q}, f_{a_l}\in F_{RS}\} $$
is a quantum hash generator. Let $s=\log{n} +\log{T}$. $G_{RS}$ generates   function
 $\psi_{G_{RS}} :(\field_q)^k\to (\Hilbert^2)^{\otimes s}$
 for a word $w\in (\field_q)^k$ by the rule
%
 %
  \begin{equation}\label{good-rs}
  \ket{\psi_{G_{RS}}(w)} = \frac{1}{\sqrt{n T} } \sum_{l=1, j=1}^{n, T} \ex{i\frac{2\pi g_{jl}(w))}{q}}\ket{lj}
  \end{equation}
 here $\ket{lj}$ denotes a basis quantum state, where $lj$ is treated as a concatenation of the binary representations of $l$ and $j$.
 \begin{property}
Let $q$ be a prime power and let $2\le k < n\le  q$. Then for arbitrary $\delta\in (0,1)$ the function $\psi_{G_{RS}}$  is an $(\epsilon,\Delta)$-resistant  $(q^k;s)$ quantum hash  function, where $\epsilon\le (q\log q)/q^k$, $\Delta\le \frac{k-1}{n}+\delta$, and  $s\le \log {(q\log q)} +2\log1/\delta +4$.
\end{property}
Let $c>1$. If we select  $n=ck$, then  $\Delta < 1/c +\delta$  and  according to  Theorem  \ref{lb} there exist constants $c_1(\Delta)$,  $c_2(\Delta)$  such that
  $\log {(q\log q)} - c_1(\Delta) \le s\le \log {(q\log q)} + c_2(\Delta)$.
Thus,  Reed Solomon codes provide balanced parameters for resistance values $\epsilon$, $\Delta$  and  for a number $s$ of qubits  for hash function $\psi_{RS}$.

\section{Presenting Quantum Hash States via Coherent States}
Written in the form given in  (\ref{good-h}) and (\ref{good-rs}), the hash states $\ket{\psi_{H_{\delta,q}}(w)}\in ({\cal H}^2)^{\otimes s}$, $w\in\field_q$, and $\ket{\psi_{RS}(w)} \in ({\cal H}^2)^{\otimes s}$, $w\in (\field_q)^k$,  need high degree of entanglement between $s$ qubits which is hard for the current technology. Papers \cite{al1_2014,al2014,aa2015} consider  the idea of presenting quantum  fingerprinting states via coherent states and developed  signature constructions based on such coherent states.

Following idea from \cite{al1_2014,al2014}, we map the hash state $\ket{\psi_{H_{\delta,q}}(w)}\in ({\cal H}^2)^{\otimes s}$ for $w\in\field_q$ to a coherent state as follows. For short we let $H_{\delta,q}=H$ in the rest of the section. Let $T=2^s$.
First, we define {\em hash mode} ($H$-hash mode) $a_{H,w}$  as
\[
a_{H,w} = \frac 1 {\sqrt T} \sum_{j=1}^T \ex{i\frac{2\pi h_j(w)}{q}} b_j,
\]
where $b_j\in \{b_1,\dots, b_T\}$ is the annihilation operator of the $j$th optical mode. Hash state is a single-photon state in the hash mode: $\ket{\psi_{H}(w)} = a_{H,w} \ket{0}$.

Next, we define {\em coherent hash state}  as $\ket{\alpha,\psi_{H}(w)}  = D_{H,w}(\alpha)\ket{0}$, with parameter $\alpha$,  where $D_{H,w}(\alpha) = \ex{\alpha a_{H,w}^\dagger - \alpha^* a_{H,w}}$ is the displacement operator. According to \cite{al2014} we have that the state $\ket{\psi_{H}(w)}$  is mapped to $\ket{\alpha,\psi_{H}(w)}$:

\[
\ket{\psi_{H}(w)} \quad \to \quad
\ket{\alpha,\psi_{H}(w) } = \bigotimes_{j=1}^T\ket{ \ex{i\frac{2\pi h_j(w)}{q}}\frac{\alpha}{\sqrt{T}}}_j,
\]
where $\ket{ \ex{i\frac{2\pi h_j(w)}{q}}\frac{\alpha}{\sqrt{T}}}_j$ is a coherent state with amplitude $\frac{\alpha}{\sqrt T}$ in the $j$th mode.

Similarly one can  map  the hash state  $\ket{\psi_{RS}(w)} \in ({\cal H}^2)^{\otimes s}$ with $w\in (\field_q)^k$ to a coherent state.
 
 In the next paper we will present a variants of quantum signature schemes based on quantum hash functions different from quantum fingerprinting function.



\end{document}